\documentclass[12pt]{article}
\usepackage{amssymb}
\usepackage{amsmath}
\usepackage{amsfonts}
\usepackage{anysize}
\marginsize{2cm}{2cm}{2.5cm}{3cm}
\date{}

\usepackage{graphicx}

\usepackage{hyperref}      

\begin{document}
\title{\LARGE\bf Manifestations of the onset of chaos in condensed matter and complex systems}

\author{Carlos Velarde\textsuperscript{1}, Alberto Robledo\textsuperscript{2}\\
\footnotesize 1. Instituto de Investigaciones en Matem\'aticas Aplicadas y en Sistemas,\\%
\footnotesize    Universidad Nacional Aut\'onoma de M\'exico, Mexico\\
\footnotesize 2. Instituto de F\'{i}sica y Centro de Ciencias de la Complejidad\\
  \footnotesize     Universidad Nacional Aut\'onoma de M\'exico, Mexico.
}
%
\maketitle
\abstract{We review the occurrence of the patterns of the onset of chaos in low-dimensional nonlinear dissipative systems in leading topics of condensed matter physics and complex systems of various disciplines. We consider the dynamics associated with the attractors at period-doubling accumulation points and at tangent bifurcations to describe features of glassy dynamics, critical fluctuations and localization transitions. We recall that trajectories pertaining to the routes to chaos form families of time series that are readily transformed into networks via the Horizontal Visibility algorithm, and this in turn facilitates establish connections between entropy and Renormalization Group properties. We discretize the replicator equation of game theory to observe the onset of chaos in familiar social dilemmas, and also to mimic the evolution of high-dimensional ecological models. We describe an analytical framework of nonlinear mappings that reproduce rank distributions of large classes of data (including Zipf's law). We extend the discussion to point out a common circumstance of drastic contraction of configuration space driven by the attractors of these mappings. We mention the relation of generalized entropy expressions with the dynamics along and at the period doubling, intermittency and quasi-periodic routes to chaos. Finally, we refer to additional natural phenomena in complex systems where these conditions may manifest.} 

\section{Introduction}
\label{intro}
The routes to chaos, period doubling, intermittency and quasi periodicity, in low dimensional dissipative systems are now classic topics in nonlinear dynamics \cite{schuster1,hilborn1,beck1}. These routes are conveniently reproduced by archetypal nonlinear iterated maps, such as the quadratic logistic and cubic circle maps \cite{schuster1,hilborn1,beck1}, and also by their extensions to general nonlinearity $z>1$ \cite{beck1,capel1,delbourgo1}.  The fascinating, often intricate, transformation of regular, periodic, motion into that of irregular, chaotic, change has attracted much attention over the years, and prompted interest about the occurrence in observable phenomena in the physical sciences and elsewhere of these basically mathematical descriptions. 

Certainly, the initial identification of manifestations of the mentioned routes and their sharp onset of chaos in various apparently independent fields of study has been sometimes unplanned, or by noticing analogies during the analysis of model systems meant to represent physical or complex system situations. But also sometimes the inverse process takes place, the consideration of some phenomenon prompts the definition of a nonlinear dynamical map, as is the early case of weak turbulence and the Pomeau-Manneville map \cite{maneville1}. As it can be observed the six instances described below tell each a different story about the discovery or development of the links between these problems or topics and nonlinear dynamics associated with the onset of chaos. But, with hindsight, a general rationalization can be advanced. As mentioned throughout and in the closing section, reduction of many degrees of freedom into a few effective degrees can take place on special circumstances or via some approximation. In addition to this there is always a contraction of phase space that the attractors of dissipative systems exert \cite{pnas1,heliyon1}. These two conditions apply in the analogies we make reference here.      

It is useful to mention briefly the sets of properties associated with the attractors involved in the sharp onset of chaos. There are those of the dynamics inside the attractors and those of the dynamics towards the attractors. These properties have been characterized in detail for the period doubling route. The organization of trajectories and the sensitivity to initial conditions are described in \cite{robledo1}, while the features of the approach of an ensemble of trajectories to this attractor is explained in \cite{robledo2}. See Fig. 1. The dynamics inside quasiperiodic attractors has also been determined by considering the so-called golden route to chaos in the circle map \cite{robledo3}. The patterns formed by trajectories and the sensitivity in this case were found to share a general structure with period doubling, but with somewhat more involved details \cite{robledo3}. On the contrary, the attractor-repellor set at the tangent bifurcation is a simpler finite set of positions with periodic dynamics and the noteworthy properties of trajectories and sensitivity correspond to the dynamics towards the attractor or away from the repellor \cite{baldovin1}.
 
The contents in this review are the following: In the next Section 2 reference is made to the analogy of the dynamics associated with the bifurcation gap in noise-perturbed period-doubling onset of chaos with glassy dynamics as observed in molecular systems. In Section 3 a description is provided for the connection between the dynamics at the tangent bifurcation and the shape and lifetimes of dominant clusters or fluctuations at critical states. In Section 4 an account is given for the equivalence between the properties of an iterated map that displays tangent bifurcations and the wave propagation properties of arrays of scatterers. Section 5 refers to the transformation into networks via the Horizontal Visibility algorithm of the families of trajectories that take place along the routes to chaos. Section 6 contains a report on the nonlinear dynamical map that displays typical routes to chaos that results from a discretized replicator equation for symmetric social games. Section 7 illustrates the iterated maps close to tangency that replicate exactly different classes of rank distributions, including Zipf's law. Finally in Section 8 there is a short summary of the topics covered and mention of some prospective additional instances of manifestations of the onset of chaos. See Ref. \cite{robentropy1} for an earlier review of some of the topics addressed here. As a difference with the presentation here, Ref. \cite{robentropy1} contains technical details and descriptions and covers topics such as: generalized statistical-mechanical structures in the dynamics along the routes to chaos, sensitivity fluctuations and their infinite family of generalized Pesin identities, development of dynamical hierarchies with modular organization, and limit distributions of sums of deterministic variables made of positions of map trajectories.

\section{Glassy dynamics from noise-induced bifurcation gap}
\label{sec:4}

It has been shown \cite{robglass1,robglass1b,robglass1c} that the dynamics at the period-doubling onset of chaos when perturbed by noise displays properties analogous to those that describe glass formation in condensed matter physics.  These properties include the occurrence of two-step relaxation, aging with its characteristic scaling property, and diminished diffusion and arrest \cite{debenedetti1,debenedetti2}. The gradual development of glassy properties that takes place with decreasing noise amplitude in ensemble- and time-averaged correlations and in diffusivity have been detailed \cite{robglass1,robglass1b,robglass1c}. See Fig. 2.

Like other dissipative chaotic dynamics, that associated with the chaotic-band attractors in unimodal maps (one-dimensional maps with one extremum) possess ergodic and mixing properties and therefore accept statistical-mechanical descriptions. At the transition out of chaos of this families of attractors, the band-splitting accumulation points that mirror the period-doubling accumulation points, ergodicity and mixing breaks down. A practical alternative to observe the decline and failure of ergodicity and mixing is to consider the noise perturbation of this accumulation point attractor, or Feigenbaum point, for which the noise amplitude is a convenient tuning parameter. As it has been known for already a long time \cite{schuster1,crutchfield1} noise produces a bifurcation gap in both period-doubling and band-splitting cascades putting an upper bound on periods and numbers of bands. These upper bounds are noise-amplitude dependent and display scaling properties \cite{schuster1,crutchfield1}. The similarity of the properties of the bifurcation gap with those of glassy dynamics was recognized in Refs. \cite{robglass1,robglass1b,robglass1c}. In particular, the ergodic to non ergodic transition that takes place when the noise amplitude vanishes was argued to be analogous to the ergodicity failure at glass formation, a considerably more difficult analysis in molecular statistical-mechanical models. The unperturbed trajectories at the Feigenbaum point have been shown to exhibit perfect aging scaling \cite{robglass1,robglass1b,robglass1c}.      

The attractors of nonlinear low-dimensional maps under the effect of external noise can be used to model states in systems with many degrees of freedom. In a one-dimensional map with only one control parameter the consideration of external noise could be thought to represent the effect of many other identical maps coupled to it, like in the so-called coupled map lattices \cite{kaneko1}. That is, the addition of noise to a unimodal map, such as the logistic map, constitutes a discrete-time Langevin equation approach for coupled-map lattices. The erratic motion of a Brownian particle is usually described by the Langevin theory \cite{chaikin1}. As is well known, this method finds a way round the detailed consideration of many degrees of freedom by representing via a noise source the effect of collisions with molecules in the fluid in which the particle moves. The approach to thermal equilibrium is produced by random forces, and these are sufficient to determine dynamical correlations, diffusion, and a basic form for the fluctuation-dissipation theorem \cite{chaikin1}.

Glass formation is associated with severe impediments to access most configurations
in a system with many degrees of freedom.
 Similarly, the dynamics of ensembles of trajectories towards the attractors
in nonlinear dissipative systems constitute realizations of phase space contraction.
 When the attractors are chaotic the contraction reaches a limit in which the
 contracted space has the same dimension as the initial space.
 Chaotic attractors have ergodic and mixing properties but those at the transitions to chaos do not \cite{beck1}. The attractors at the transitions to chaos provide a natural mechanism by means of which ensembles of trajectories are forced out of almost all phase space positions and become confined into a finite or (multi)fractal set of permissible positions \cite{heliyon1}.

\section{Growth, collapse and intermittency of critical fluctuations}
\label{sec:5}

The properties of a large, statistically dominant, critical fluctuation, or cluster, have been determined \cite{athens1,athens2,athens3,athens4,robcrit1,robcrit1b} and a description has been obtained for the shape of its order parameter $\phi$ in a region of size $R$. These studies \cite{athens1,athens2,athens3,athens4,robcrit1,robcrit1b} make use the statistical-mechanical method of a coarse-grained free energy, as it is the Landau-Ginzburg-Wilson (LGW) continuous spin model Hamiltonian, at the critical temperature and with zero external field. The $\phi$-shape of the statistically dominant cluster of radius $R$ has been shown to have a fractal configuration \cite{athens1,athens2}, while its amplitude $\phi$ grows in time and eventually collapses when an instability is reached \cite{athens3,athens4}. Interestingly, a nonlinear map with tangency and feedback features describes this process, such that the time evolution of the cluster is given in the nonlinear system as a laminar episode of intermittent dynamics \cite{athens3,athens4,robcrit1,robcrit1b}.

Recently \cite{robcrit1c}, the relationship that exists between such critical clusters in thermal systems and intermittency near the onset of chaos in low-dimensional systems has been re-examined. Use has been made of the density functional method for inhomogeneous systems and of the renormalization group (RG) method in nonlinear dynamics to provide a more accurate and formal account of the subject. The description of this significant correspondence includes, on the one hand, the density functional formalism, where classical and quantum mechanical analogues match the procedure for one-dimensional clusters, and, on the other, the RG fixed-point map of functional compositions that captures the basic dynamical behavior. Implicit in these studies there is a correspondence between the high-dimensional (degrees of freedom) phenomenon and low-dimensional dynamics.

A tractable simplification of the study of a general fluctuation or cluster of the order parameter $\phi$ is that of a one-dimensional thermal system undergoing a second order phase transition. The choice of a one-dimensional system provides access to a analytical spatial description of $\phi$ that still shares some of the main properties of clusters in higher spatial dimensions. As known \cite{fisher1}, criticality in one-dimensional systems requires long-range interactions. The dominant shape of the fluctuation or cluster in the large size $R$ limit is obtained via the steepest-descent approximation on the partition function for the LGW Hamiltonian. This method is equivalent to the minimization of a free energy density functional and allows for explicit detail. The first variation of the free energy functional, or Euler-Lagrange (EL) equation turns out to be analogous to the description of trajectories of a classical particle. This analogy makes it possible to determine the possible types of order parameter profiles for the critical cluster. See Figs. 3a and 3b. Further, the force that the different types of $\phi$-profiles apply on its boundary determines its growth or shrinkage of the cluster. The calculation of the second variation of the free energy functional leads to an expression of
 the form similar to a Schrodinger operator  
 for a quantum particle. The eigenfunctions of
 this operator 
 correspond to perturbations of the order parameter profile while the eigenvalues indicate the stability of the profile to such perturbations.

As for the temporal properties of the cluster the collapse or the intermittent behavior of the critical cluster have been considered. The phase-portrait equation obtained (after one integration) from the EL equation can be transformed into a nonlinear iteration map that is used as the starting point for the calculation of the renormalization group (RG) fixed-point map near tangency \cite{hu1}. Perturbation of this map \cite{hu1} takes it away from tangency and this leads, on the one hand, to trajectories that describe the collapse of the cluster, and, on the other hand, to trajectories made of laminar episodes interrupted by rapid bursts, as is the intermittent dynamics associated to this type of transition to or out of chaos See Figs. 3c and 3d. The critical cluster develops during the laminar episode leading to sudden breakdown that is followed by the growth of a new cluster. The power spectrum for this dynamical behavior is known \cite{procaccia1} to be of the $1/f$ noise kind characteristic of critical fluctuations.

\section{Localization and transport near and at a tangent bifurcation}
\label{sec:6}

It has been found that the scattering properties of two related models, a layered linear periodic structure and a regular Bethe-like network (see Fig. 4a), are exactly given by those for the dynamics of nonlinear maps that exhibit tangent bifurcations (see Fig. 4b) \cite{robloc1,robloc2,robloc3,robloc4}. The presence of these maps is a consequence of the combination rule of scattering matrices when the scattering systems grow in size via consecutive replication of an element or motif. Although so far limited to the model systems studied, an equivalence has been shown between wave transport phenomena in classical wave systems, or in electronic transport through quantum systems, and the dynamical properties of low-dimensional nonlinear maps, specifically at the onset of chaos associated with intermittency of type I \cite{schuster1}. This is a remarkable property in that a system with many degrees of freedom experiences a radical reduction of these, so that its description is totally provided by only a few variables.

The transitions from localized to
 delocalized 
 states correspond to transitions from regularity to chaos \cite{robloc1,robloc2,robloc3,robloc4} and the chaotic regimes hold an ergodic property \cite{schuster1} that can be used to obtain scattering ensemble averages. In the language of the quantum case the conductance in the crystalline limit reflects the periodic or chaotic nature of the attractors (see Fig. 4c). The transition between the insulating to conducting phases can be seen as the transition along the known route to (or out of) chaos at a tangent bifurcation \cite{schuster1}. While the conductance displays an exponential decay with size in the evolution towards the crystalline limit, it obeys instead a power law decay at the transition. A similar behavior can be found for a locally periodic structure where the wave function also decays exponentially within a regular band (or regular attractor) and a power law decay with system size at the mobility edge (or onset of chaos). A complete transcription from the nonlinear dynamical language to that of the scattering process can be made, such that, for instance, the vanishing of the Lyapunov exponent at the transitions to chaos corresponds to the divergence of the localization length at the gap to band boundaries.
 
It has also been shown \cite{robloc2} that the localized and delocalized states of large regular networks of scatterers can be described generically by negative and zero Lyapunov exponents, respectively, of M\"obius group transformations that describe the evolution of the scattering matrix as the system size grows. See Fig. 4d. This development constitutes a rare simplification in which there is a large reduction of degrees of freedom: a system composed by many scatterers is described by a low-dimensional map. This property is reminiscent of the drastic reduction in state variables displayed by large arrays of coupled limit-cycle oscillators, for which their macroscopic time evolution has been shown \cite{strogatz1} to be governed by underlying low-dimensional nonlinear maps of the M\"obius form. The parallelism between the transport properties of the networks studied in \cite{robloc2} and the dynamical properties of arrays of oscillators is made more visible by noticing that in both problems the variables of interest are similar: phase shift of the scattered states and phase change with time of coupled oscillators, both determined by the action of the M\"obius group. In the scattering problem we arrive at the basic nonlinear map by first constructing a family of self-similar networks, and then relating the scattering matrices that represent two members of the family. In the case of coupled oscillators the puzzle of the drastic reduction of variables finds a rationale in the identification of the role of the M\"obius group in the temporal evolution of the system. Notice that the particulars of the scattering potentials appear only through the coefficients of the M\"obius transformation while the structure of the network gives the transformation its general form \cite{robloc2}.

\section{A network view of the three routes to chaos}
\label{sec:2}

The dynamical properties of the three routes to chaos in one-dimensional nonlinear maps, period doubling, intermittency and quasi periodicity, has been converted into network properties \cite{robtolo1,robtolo2,robtolo3,robtolo4,robtolo5} via the horizontal visibility (HV) algorithm that transforms time series into graphs (see Fig. 5a) \cite{tolo1,tolo2}. An important feature of HV graphs is that each one of them represents a large number of map trajectories, that is, many time series lead to the same HV graph, and each of them captures significant characteristics of a class of trajectories. In our case studies, the three routes to chaos, each HV graph represents a distinct family of attractors. Sets of graphs have been obtained for the period-doubling cascade and the intermittent trajectories near tangent bifurcations in unimodal maps, as well as the quasiperiodic routes to chaos in circle maps. See Figs. 5b, 5c and 5d. These sets of graphs are independent of the details of the unimodal or circle maps including their nonlinearity.

The families of HV graphs obtained for the three routes to chaos allow for the calculation of simple closed expressions for their degree $k$ distribution $p(k)$, and through them it is possible to obtain exact quantitative access to their entropy. Interestingly, the HV graphs are entropy extremes. Also, it is straightforward to define Renormalization Group (RG) transformations for the families of HV graphs, and the RG flow diagrams and trivial and nontrivial fixed points have been determined for the three routes to chaos, as well as for the HV graphs that correspond to chaotic attractors. This has led to a comprehensive description of their self-similar properties in which the important nontrivial fixed-point graphs correspond to those that represent the transitions to chaos. The HV graph families have been ordered along RG flows and their limiting fixed points have been determined.

An important question pointed out some time ago \cite{robRG1}, is whether there exists a connection between the extremal properties of entropy expressions and the renormalization group approach. Namely, that the fixed points of RG flows can be obtained through a process of entropy optimization, adding to the RG approach a variational quality. The families of HV graphs obtained for the three routes to chaos offer a valuable opportunity to examine this issue. The answer provided by HV graphs to the question posed above is clearly in the affirmative \cite{robtolo1,robtolo2,robtolo3,robtolo4,robtolo5}. In these cases we observe the familiar picture of the RG treatment of a model phase transition, two trivial fixed points that represent disordered and ordered, or high and low temperature, phases, and a nontrivial fixed point with scale-invariant properties that represents the critical point. There is only one relevant variable, the difference in the map control parameter value from that at the transition to chaos, that is necessary to vanish to enable the RG transformation to access the nontrivial fixed point.

There has been also focus on networks generated by unimodal maps at their period-doubling accumulation points with regards to their fluctuations in connectivity as their size grows. First, it was found \cite{robtolo6} that the expansion of connectivity fluctuations admits the definition of a graph-theoretical Lyapunov exponent. Second, the expansion rate of trajectories in the original accumulation point attractor translates into a generalized entropy that surprisingly coincides with the spectrum of generalized graph-theoretical Lyapunov exponents \cite{robtolo6}. This result indicates that Pesin-like identities valid at the onset of chaos \cite{robentropy1} are also found in complex networks that possess certain scaling properties \cite{robtolo6}. The study of connectivity fluctuations has been extended to the case of the quasiperiodic route to chaos in circle maps \cite{robtolo7} and similar results have been obtained. See also \cite{robtolo5}.

\section{Chaos in discrete time game theory}
\label{sec:3}

The study of a discrete-time version of the replicator equation for two-strategy games has revealed that stationary properties differ considerably from those of continuous time for sufficiently large values of the pay-off parameters \cite{robvilone1}. Two-strategy symmetric games become a one-dimensional map with two control parameters. The familiar period doubling and chaotic band-splitting attractor cascades of unimodal maps have been observed, but in some cases more elaborate variations appear due to bimodality. The discretized version of the replicator dynamics shows that there are large regions of parameter space, i.e. temptation to defect ($T$) and risk in cooperation ($S$), $(T, S)$, of the typical social dilemmas with behavior largely different from the continuous one. Indeed, beyond the restricted region that has been commonly analyzed before, it has been observed that as the parameters $T$ and $S$ increase, periodic ($P>1$) and chaotic dynamics occurs (see Fig. 6a) \cite{robvilone1}. This is an important difference with respect to the continuous time dynamics, for which only fixed-point solutions ($P=1$) are found. Periodic solutions imply that cooperation ($C$) and defection ($D$) strategies vary periodically, the state of the system being generally mixed, containing both types of individuals. The occurrence of chaotic behavior, including intermittency, adds unpredictability to the features of the system, in complete contrast with the predictive character of the continuous case. The results may find relevance for real life applications, in many cases involving bacteria, animals or humans which are well described by social dilemmas that do not have well-determined payoffs, and often take place in discrete time. The appearance of periodic orbits and chaos is generic and will affect other strategic interactions when described by a discretized equation. Unimodal maps (obtained, e.g., when $A \equiv T=S<0$) show a rich self-affine structure of chaotic-band attractors interspersed by periodic windows within which multiple period-doubling and band-splitting cascades take place. Bimodal maps (obtained, e.g., when $A>0$) display more elaborate attractor structures (see Fig. 6b) \cite{robvilone1}.

Recently \cite{robjensen1}, it has been found that the same discrete replicator equation appears in a radical simplification of a game-theoretic adaptation of the Tangled Nature (TaNa) model \cite{robjensen2} for evolutionary dynamics of ecological systems. The original game-theoretic version of the model for the dynamics of ecosystems is described by a large set of coupled replicator equations that also includes stochastic mutations \cite{jensen1}. This model exhibits macroscopic non-stationary intermittent evolution similar to that in the TaNa model, and, interestingly, its discrete time version that operates in the limit of many strategies constitutes a coupled map lattice (CML). Subsequently, a major simplification of these CML replicator-mutation equations was considered to reduce the model into a one-dimensional nonlinear map \cite{robjensen1}. With this approximation the possible connection was examined between the macroscopic intermittent behaviors of the above-mentioned high-dimensional models with the low dimensional established sources of intermittency, such as the tangent bifurcation \cite{schuster1} with known $1/f$ noise spectra \cite{procaccia1}. As a result, the many-strategy game-theoretic problem has been reduced to a classic version of two strategies, where one of them represents a selected agent or species and the other groups all the others. One advantage in using this approximation is that a one-dimensional nonlinear dynamical model can be constructed \cite{robjensen2} such that its time evolution consists of successive tangent bifurcations that generate patterns resembling those of the full TaNa model in macroscopic scales. See Figs. 6c and 6d. The parameters in the model are based on identified mechanisms that control the duration of the basic quasi-stable event generated by a local mean-field map derived from the TaNa model \cite{robjensen2}. 

\section{Nonlinear map analogues of rank distributions}
\label{sec:7}

A precise analogy has been developed between stochastic and deterministic variable approaches to represent ranked data. And a statistical-mechanical interpretation of the expressions obtained has been pointed out as the possible explanation of the generality of their functional shapes \cite{robzipf1a,robzipf1b,robzipf1c,robzipf1d}. Size-rank distributions $N(k)$ from real data sets are reproduced by forthright considerations based on the assumed knowledge of the parent or source probability distribution $P(N)$ that generates samples of random variable values similar to real data. Then, the selection of various functional expressions for $P(N)$: power law, exponential, Gaussian, etc., are seen to produce different classes of distributions $N(k)$ for which there are real data examples \cite{robzipf1d}. It has been found \cite{robzipf1d} that all of these types of functions $N(k)$ can also be obtained from deterministic dynamical systems. These correspond to one-dimensional nonlinear iterated maps close to tangency with the identity line and in some cases near a tangent bifurcation. The trajectories of these maps were proved to be exact analogues of the stochastic expressions for $N(k)$. The map trajectories were in all cases found to occur for small Lyapunov exponent, therefore near a transition between regularity and chaos. The explicit examples range from exponential to logarithmic behavior, including Zipf's law. Adoption of the nonlinear map as the formalism central concept appears to be a useful viewpoint, as variation of its few parameters, that modify its tangency property, translate into the different classes for $N(k)$ \cite{robzipf1d}. See Fig. 7.

The relationship between frequency and magnitude ranked data has also been examined recently \cite{robzipf1c}. It was found that the size and frequency rank distributions are functional inverses of each other. This relationship was established both in terms of the assumed parent or source probability distribution that generates the data samples, and also in terms of the analog deterministic nonlinear iterated map that reproduces them. For the particular case of hyperbolic decay with rank the distributions are identical, that is, the classical Zipf plot, a pure power law. But their difference is largest when one displays logarithmic decay and its counterpart shows the inverse exponential decay, as it is the case of Benford law, or vice versa. For all intermediate decay rates generic differences appear not only between the power-law exponents for the central part rank decline but also for small and large rank \cite{robzipf1c}. To corroborate the findings data were selected for earthquakes and forest fire areas as these can be ranked either way, according to numbers of occurrences or sizes \cite{robzipf1c}. Also data that obeys Benford law was considered and agreement was found \cite{robzipf1c}. 

It has been shown \cite{pnas1,heliyon1} that size-rank distributions with a power-law decay are associated with generalized entropy expressions of the Tsallis $q$-deformed type. These expressions are derived from a maximum entropy principle with the use of two different constraints, and the resulting duality of entropy indexes $\alpha$ and $\alpha'=2-\alpha$ has been considered to carry physically relevant information. Whereas the value of the index $\alpha$ fixes the distribution's power-law exponent, that for the dual index $\alpha'$ fixes the extensivity of the generalized entropy \cite{pnas1,heliyon1}. It has also been argued that dual entropy expressions apply naturally to statistical-mechanical systems that experience a radical contraction of their configuration space. The entropic index $\alpha > 1$ describes the contraction process, while the dual index $\alpha' = 2 -\alpha < 1 $ defines the contraction dimension at which extensivity is restored. This circumstance has been verified \cite{heliyon1} along the three routes to chaos where the attractors at the transitions between regular and chaotic behavior drive phase-space contraction for ensembles of trajectories. This condition extends to properties of systems that find descriptions in terms of nonlinear maps.

\section{Summary \& prospective}
\label{sec:8}

We have reviewed recent studies in half a dozen topics that involve the well-known routes to chaos in low-dimensional nonlinear systems as displayed by the classic logistic and circle maps. The topics refer to sharp collective phenomena (glass formation, critical points and localization transitions) in condensed matter physics, and to objects of study (networks), general techniques (game theory), and widespread observations (rank distributions) in complex systems with large numbers of degrees of freedom. Mainly, these topics have developed independently from the perspective and concepts of nonlinear dynamics, and here we have called attention to analogies and links that exist between these different fields.  

We have mentioned that the connections discussed between low-dimensional nonlinear dynamics and high dimensional many body or agent systems involve a drastic reduction of variables, together with the further reduction that results from the action on trajectories by the attractors of the nonlinear dissipative mappings, particularly at the transitions to chaos. We have mentioned the occurrence of generalized entropy expressions in the studies reviewed here, as this is evocative of the generality of the few-variable formalism of thermodynamics of equilibrium states in ordinary statistical-mechanical systems.

Other possible phenomena in complex systems where the characteristics of the onset of chaos may appear in ways similar to those illustrated here are: (1) Protein folding \cite{protein1}, traffic jams \cite{traffic1} and hill slope evolution \cite{hillslope1} are processes that display glassy dynamics and could be modelled via the bifurcation gap. (2) Critical fluctuations in many kinds of systems, like, to give one example, in spatial complex networks \cite{spatialnetworks1}. (3) Localization-delocalization of light, sound or other propagating particles and waves through scattering media \cite{scattering1}, including motion of insects and related diffusion phenomena in biological sciences \cite{insects1}. Also: (4) It may be fruitful to convert time series from various sources that are thought to display features related to the onset of chaos into Horizontal Visibility networks, like, for instance, cardiovascular time series \cite{cardiovascular1}. (5) The replicator dynamics of evolutionary game theory is a tool that has found many applications in complex systems \cite{gametheory1}, as we have described its discrete time version brings the dynamical properties of nonlinear iterated mappings within reach of these endeavors. (6) Ranking is widely used to render data of numerous phenomena in many different fields of social and natural sciences. We have suggested that ranking distributions and their nonlinear dynamical analogues indicate an underlying statistical-mechanical structure with universal qualities \cite{robzipf1c,robzipf1d}.


\newpage
\def\captionFigI{%
\caption{Some dynamical properties associated with period doubling.
  (a) Trajectory within the attractor at the period-doubling accumulation point \cite{robledo1,robentropy1}.
  (b) Trajectories expansion rate at the period-doubling accumulation point \cite{robledo1,robentropy1}.
  (c) Sequential gap formation of an ensemble of trajectories en route to the
  attractor at the period-doubling accumulation point \cite{robledo2,robentropy1}.
  (d) Flight times $t_f$ (number of iterations) for trajectories to reach a period 8 attractor \cite{robledo2,robentropy1}.}
}

\def\captionFigII{%
\caption{Dynamical properties of the noise-induced bifurcation gap similar to those of glassy dynamics.
  (a) Families of attractors for the logistic map in the absence of noise,
  the insets show attractor broadening by noise of amplitude $\sigma=10^4$ \cite{robentropy1}.
  (b) Decay of two-time correlation for trajectories at the onset of chaos for various
  noise amplitudes $\sigma$ \cite{robglass1c}.
  (c) Aging scaling property of two-time correlations for trajectories in the absence of noise \cite{robglass1c}.
  (d) Abatement of diffusion shown by  mean-square displacement in (repeated-cell) logistic map at onset
  of chaos as noise amplitude decreases \cite{robglass1c}.}
}

\def\captionFigIII{%
\caption{Properties of a large dominant critical fluctuation.
  (a) Order parameter profile for a growing cluster before instability \cite{robcrit1c}.
  (b) Order parameter profile for a collapsing cluster \cite{robcrit1c}.
  (c) Intermittency of fluctuations via de perturbed Hu and Rudnick fixed-point map \cite{robcrit1c}.
  (d) Collapse of fluctuations via de perturbed map \cite{robcrit1c}.}
}

\def\captionFigIV{%
\caption{Localized and delocalized properties of lattice arrangements of scatterers.
  (a) A double Cayley tree lattice model where nodes are identical scatterers \cite{robloc1}.
  (b) The phase of the scattering matrix  for successive lattice sizes leads to a
  nonlinear map close to tangent bifurcations \cite{robloc1}.
  (c) Bifurcation diagram formed by period one (localized states) and chaotic
  (delocalized states) attractors \cite{robloc1}.
  (d) The model scattering matrix performs M\"obius group transformations in the
  unit complex circle as the system size grows \cite{robloc2}.}
}

\def\captionFigV{%
\caption{Families of Horizontal Visibility (HV) networks obtained along the routes to chaos.
  (a) Illustration of the HV algorithm that converts time series into a graph \cite{robtolo5,tolo1,tolo2}.
  (b) Sequence of HV graphs for consecutive period doubling attractors \cite{robtolo1,robtolo2,robtolo5}.
  (c) Farey tree for the quasi periodic route to chaos illustrating HV graph motifs and
       periodic attractor sequences for gold and silver numbers \cite{robtolo3,robtolo5}.
  (d) Illustration of the fixed-point HV graph for the tangent bifurcation of period three \cite{robtolo4,robtolo5}.}
}

\def\captionFigVI{%
\caption{Dynamical properties of the discrete-time replicator equation for $2\!\times\!2$ symmetric games.
  (a) Lyapunov exponent in control parameter plane $(S,T)$ (blue indicates negative
  value and yellow/orange positive. Inset illustrates payoff matrix and parameter location
  of familiar social games \cite{robvilone1}.
  (b) Bifurcation diagram and Lyapunov exponent along the parameter line $A \equiv S=T$ \cite{robvilone1}.
  (c) Five-times-composed replicator map at the tangent bifurcation of its period five
  window with a close view inset \cite{robjensen1}.
  (d) Trajectory for the conditions in (c) that imitates the behavioral patterns of a high-dimensional model for ecological evolution \cite{robjensen1}.}
}

\def\captionFigVII{%
\caption{Examples of real ranked data according to size (blue) and fitting obtained
  from trajectories (red) of different types of iterated maps close to tangency \cite{robzipf1d}.
  (a) Forest fire areas, ordinates in ordinary logarithmic scale \cite{robzipf1c}.
  (b) Infant mortality per country in ordinary linear scales \cite{robzipf1d}.
  (c) Fireams owned per 100 capita and per country, in ordinary logarithmic scales \cite{robzipf1d}.
  (d) Los Angeles household sizes, ordinates in ordinary logarithmic scale \cite{robzipf1d}.}
}


\begin{figure}[!h]
\resizebox{1.00\columnwidth}{!}{%
     \includegraphics{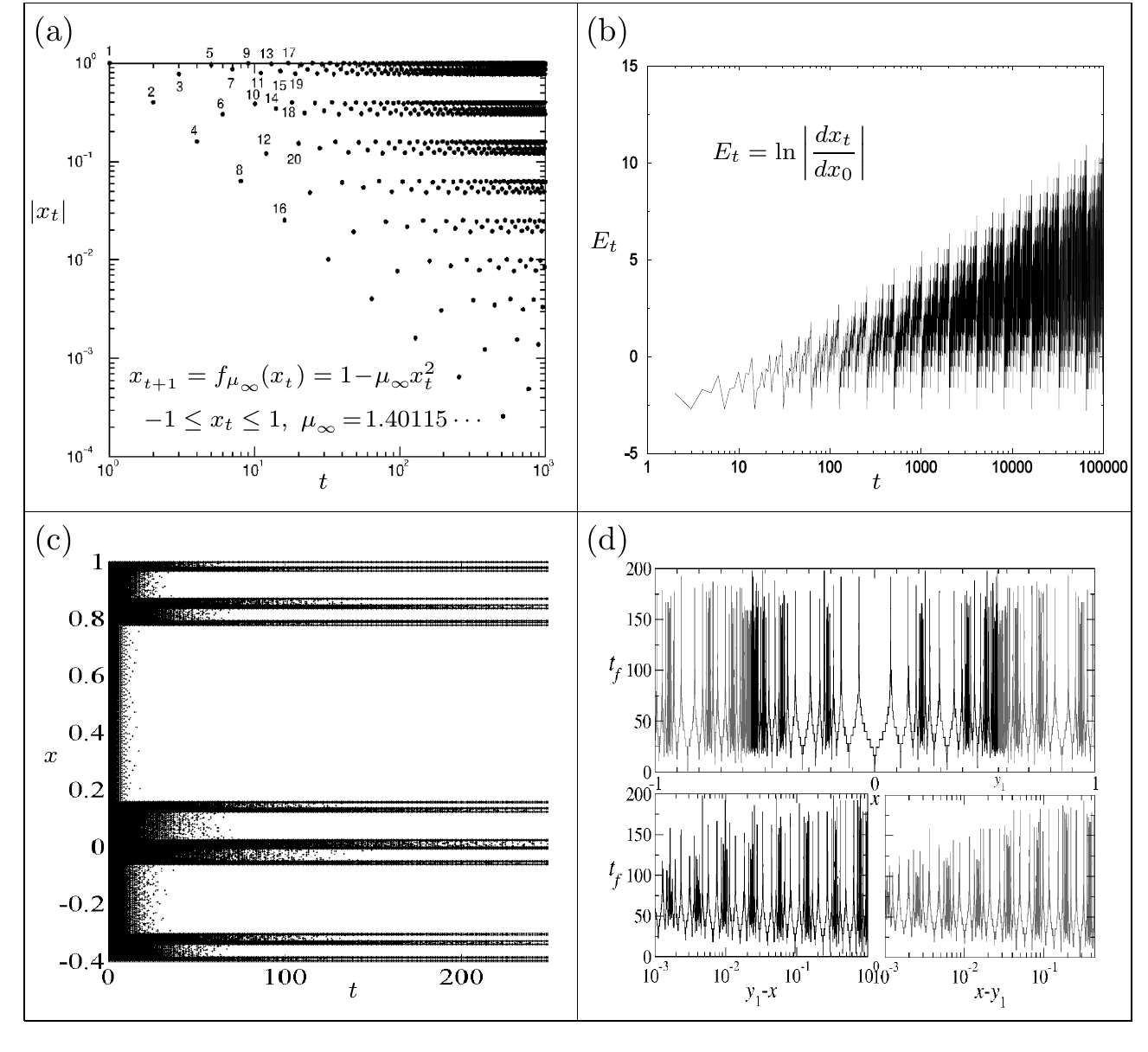} }
\captionFigI
\label{fig:Raw}
\end{figure}

\begin{figure}[!h]
\resizebox{1.00\columnwidth}{!}{%
     \includegraphics{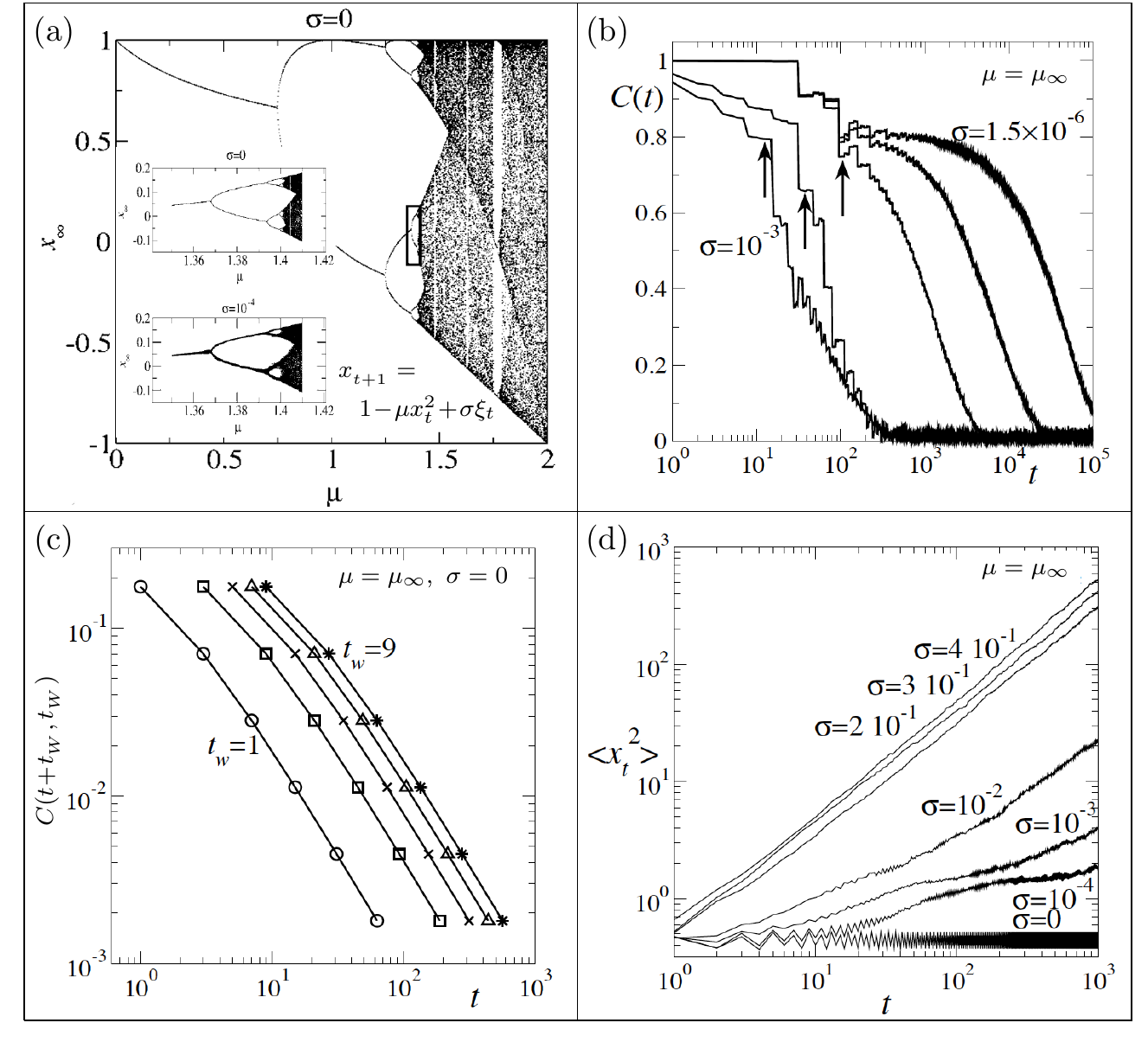} }
\captionFigII  
\label{fig:Glassy_B}
\end{figure}

\begin{figure}[!h]
\resizebox{1.00\columnwidth}{!}{%
     \includegraphics{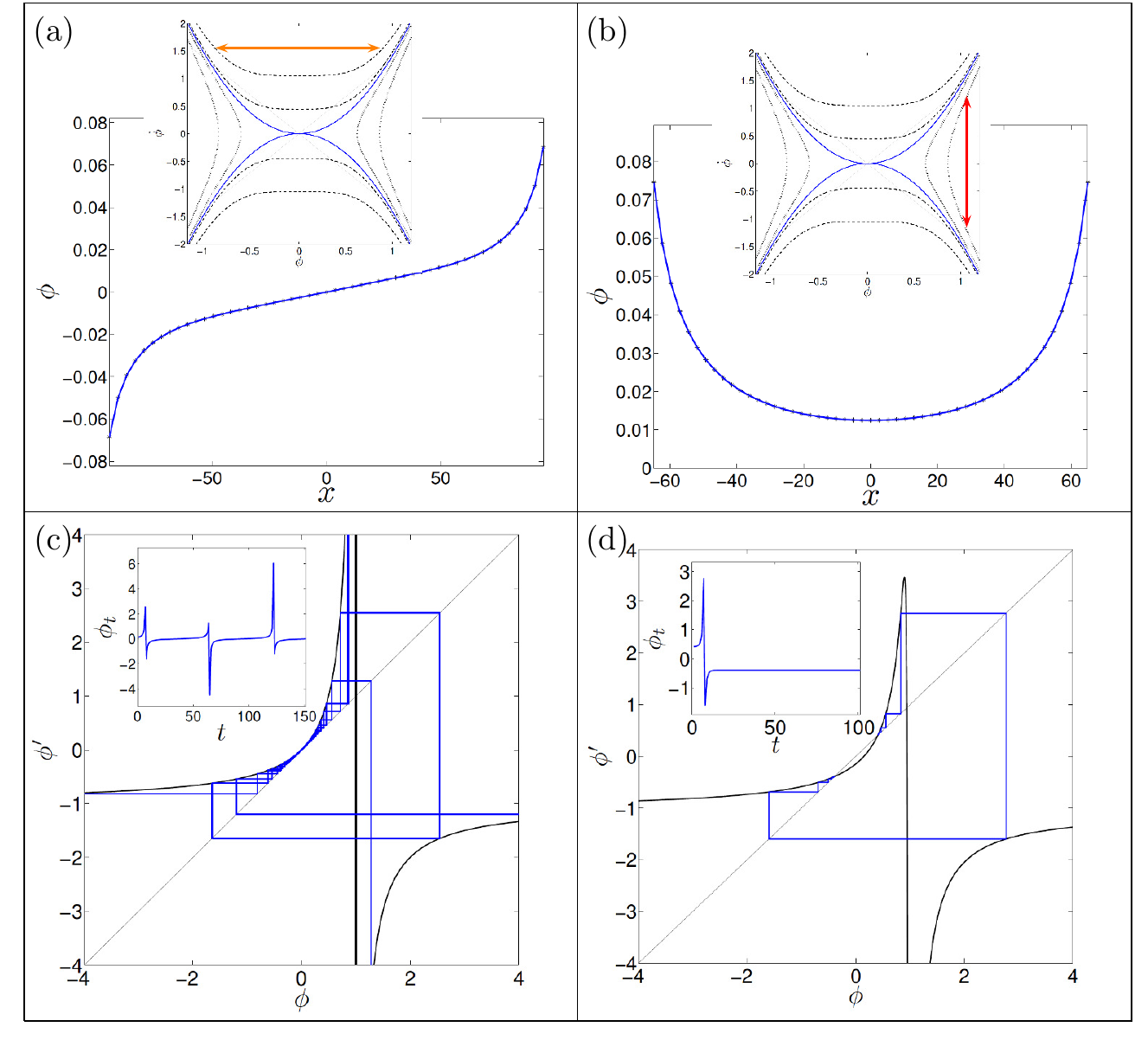} }
\captionFigIII     
\label{fig:Critical}
\end{figure}

\begin{figure}[!h]
\resizebox{1.00\columnwidth}{!}{%
     \includegraphics{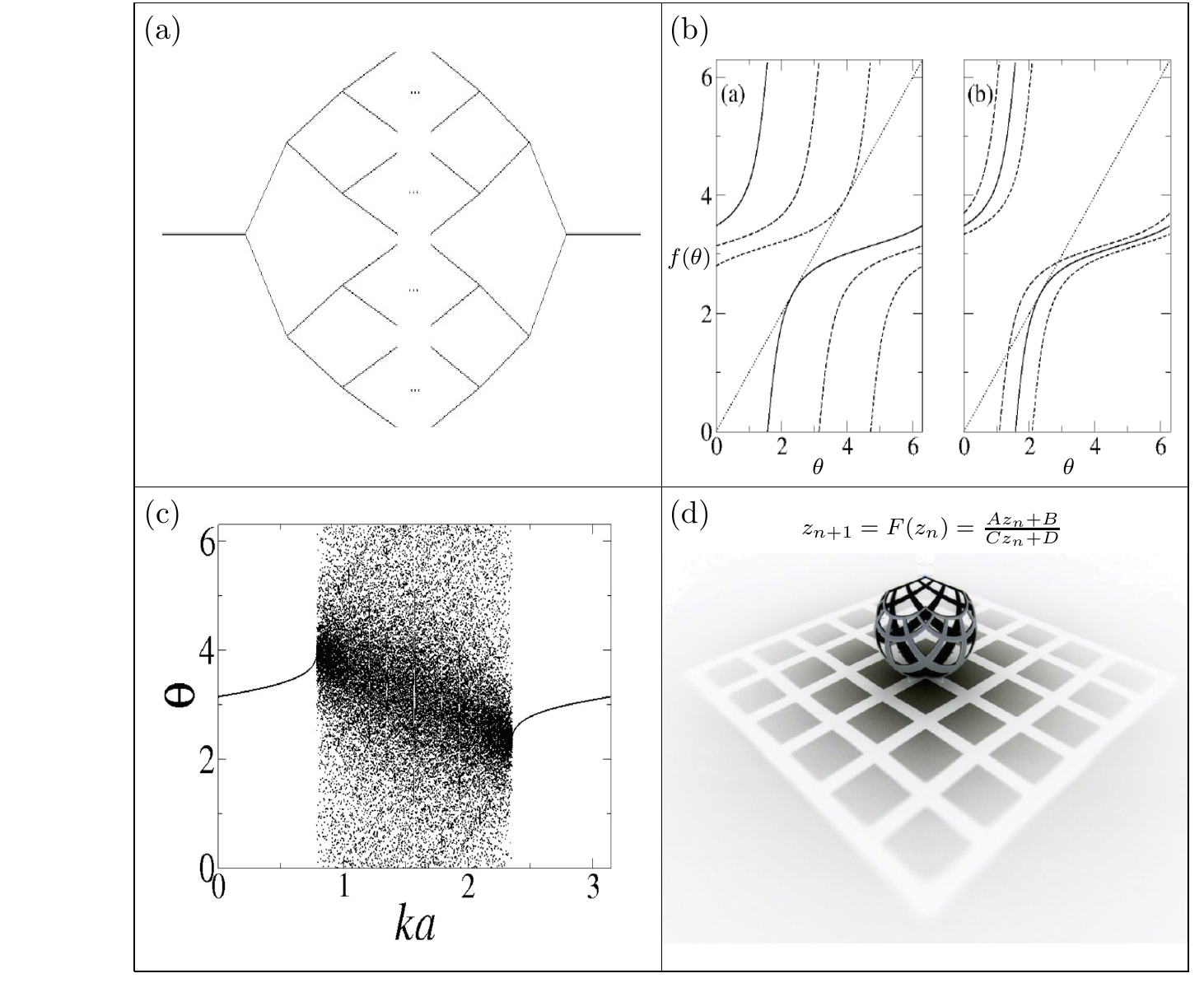} }
\captionFigIV        
\label{fig:Localization}
\end{figure}

\begin{figure}[!h]
\resizebox{1.00\columnwidth}{!}{%
     \includegraphics{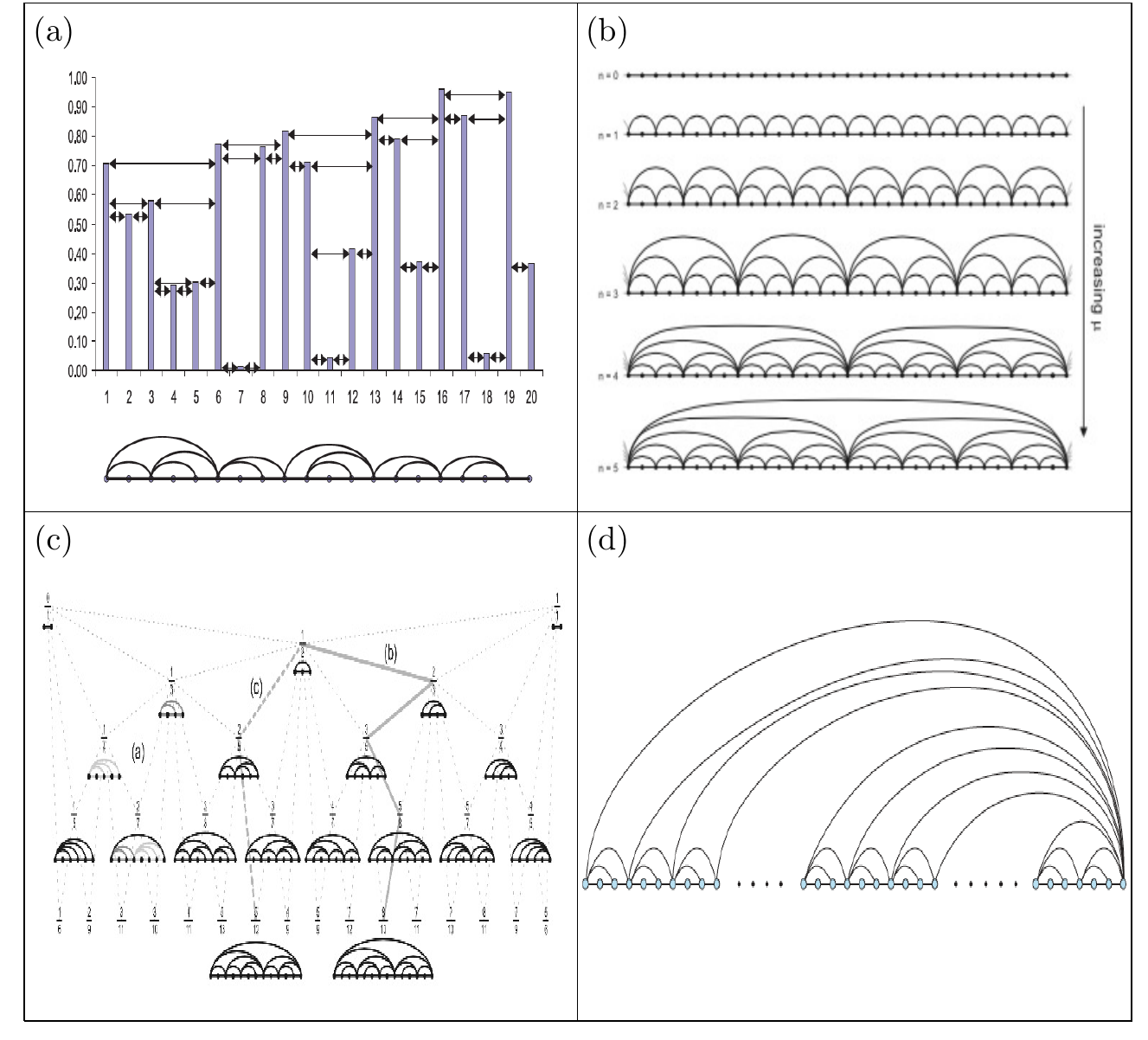} }
\captionFigV    
\label{fig:HVG}
\end{figure}

\begin{figure}[!h]
\resizebox{1.00\columnwidth}{!}{%
     \includegraphics{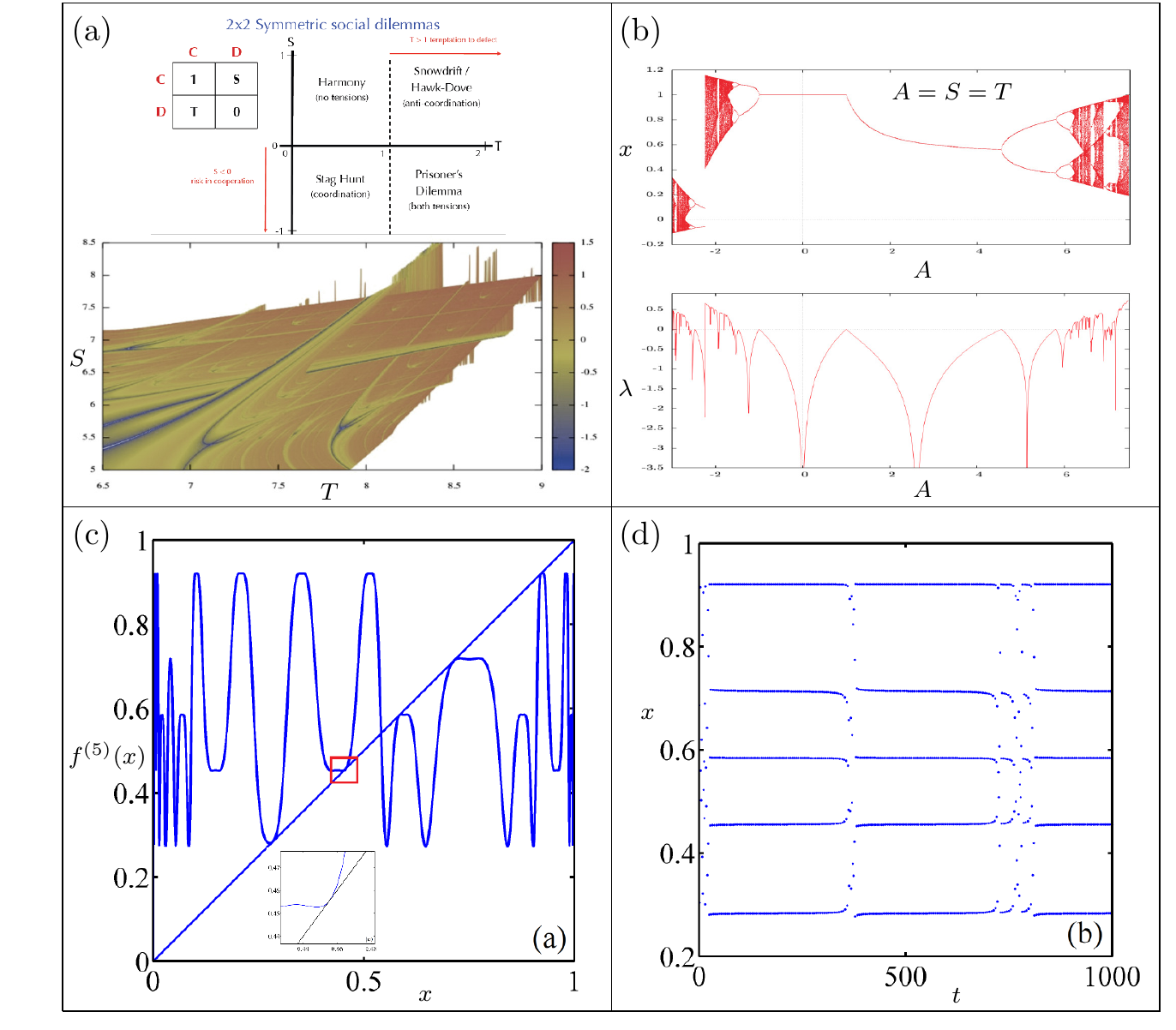} }
\captionFigVI       
\label{fig:Game}
\end{figure}

\begin{figure}[!h]
\resizebox{1.00\columnwidth}{!}{%
     \includegraphics{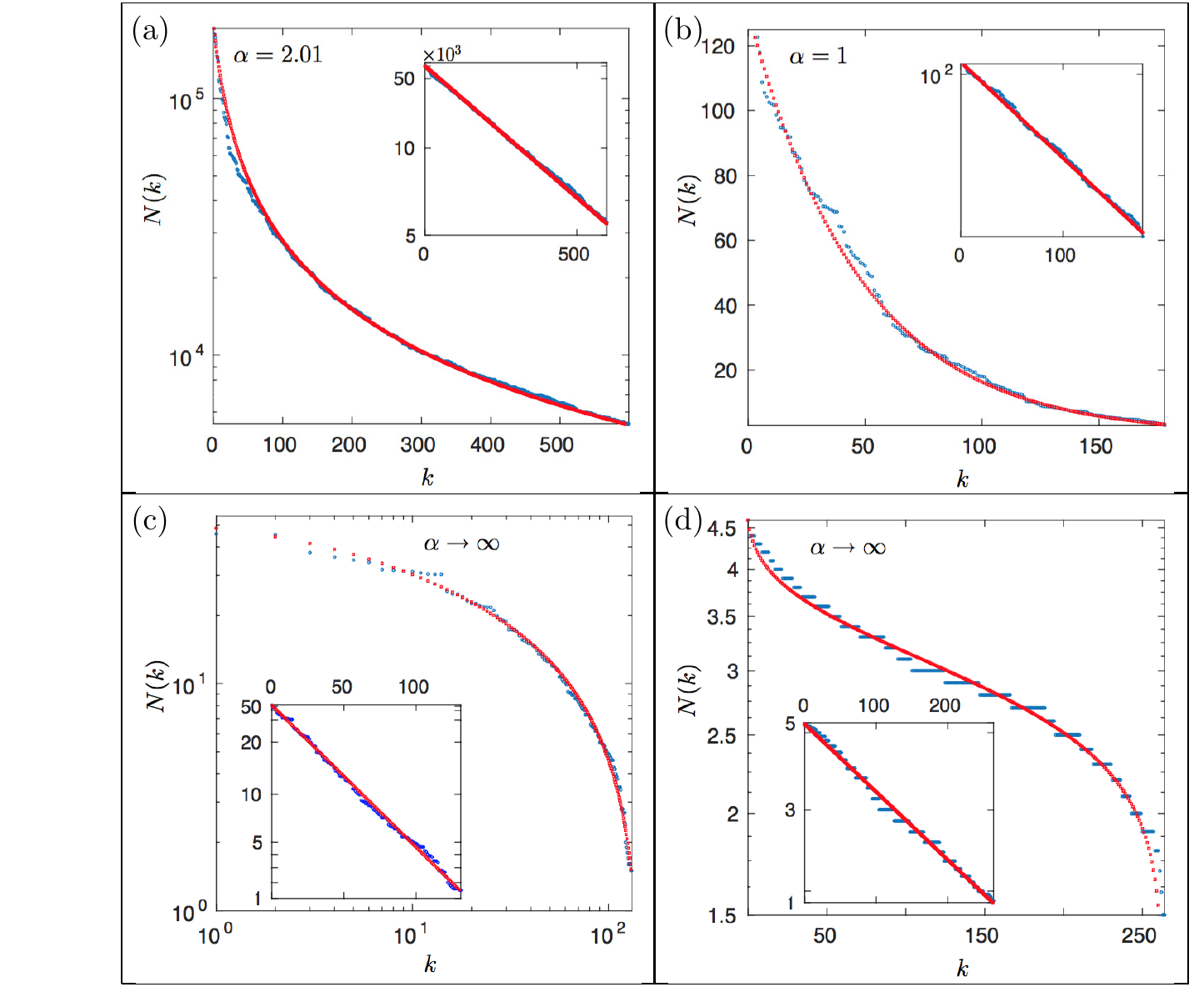} }
\captionFigVII   
\label{fig:Rank}
\end{figure}

\end{document}